\documentclass[aps,10pt,reprint,groupedaddress,superscriptaddress]{revtex4-2}

\usepackage[usenames,dvipsnames]{xcolor}
\usepackage{amssymb}
\usepackage{graphicx}
\usepackage{amsmath}
\usepackage[citecolor=blue, colorlinks=true, urlcolor=blue, linkcolor=blue]{hyperref}
\usepackage{braket}
\usepackage[compat=1.1.0]{tikz-feynman}
\usepackage[normalem]{ulem}
\definecolor{pink2}{rgb}{0.858, 0.188, 0.478}

\begin{document}
\title{Characterizing fractional topological phases of lattice bosons near the first Mott lobe}

\newcommand{\TUM}{\affiliation{Department of Physics and Institute for Advanced Study, Technical University of Munich, 85748 Garching, Germany}}
\newcommand{\MCQST}{\affiliation{Munich Center for Quantum Science and Technology (MCQST), Schellingstr. 4, 80799 M{\"u}nchen, Germany}}

\author{Julian Boesl}  \TUM \MCQST
\author{Rohit Dilip}  \TUM \MCQST
\author{Frank Pollmann} \TUM \MCQST
\author{Michael Knap} \TUM \MCQST

\begin{abstract}
The Bose-Hubbard model subjected to an effective magnetic field hosts a plethora of phases with different topological orders when tuning the chemical potential. 
Using the density matrix renormalization group method, we identify several gapped phases near the first Mott lobe at strong interactions. They are connected by a particle-hole symmetry to a variety of quantum Hall states stabilized at low fillings.
We characterize phases of both particle and hole type and identify signatures compatible with Laughlin, Moore-Read, and Bosonic Integer Quantum Hall states by calculating the quantized Hall conductance and by extracting the topological entanglement entropy. 
Furthermore, we analyze the entanglement spectrum of Laughlin states of bosonic particles and holes for a range of interaction strengths, as well as the entanglement spectrum of a Moore-Read state.
These results further corroborate the existence of topological states at high fillings, close to the first Mott lobe, as hole analogues of the respective low-filling states.
\end{abstract}

\maketitle

\section{Introduction}

Exotic quantum phases have been discovered over the past decades that defy the Landau classification of spontaneous symmetry breaking and instead exhibit topological order~\cite{Tsui_1982, Wen2017}. These phases possess emergent quasi-particles with anyonic exchange statistics and long-range quantum entanglement, which renders their experimental characterization challenging. The development of quantum simulators and quantum computers offers new opportunities to realize and characterize such novel topological phases. Topological entanglement has been measured and anyon braiding has been simulated with a superconducting qubit quantum processor~\cite{satzinger2021}, and non-local order parameters have been observed in a Rydberg atom quantum simulator~\cite{semeghini2021probing}. Recent progress also enabled the study of quantum Hall physics with ultracold atoms in optical lattices, where Floquet driving creates artificial gauge fields~\cite{Aidelsburger2013, Miyake2013, Jotzu2014, Atala_2014, Aidelsburger2015, Stuhl_2015, Mancini_2015, Flaschner2016}. Utilizing these techniques, the interaction-induced chiral propagation of excitations has been explored in the few-body limit of quantum Hall states~\cite{Tai2017, Roushan2017}. Besides that, several numerical studies predict that lattice bosons in strong effective magnetic fields realize a plethora of fractional quantum Hall states, see e.g. Refs.~\onlinecite{Sorensen_2005, Palmer_2006, Hafezi_2007, Moller2009, Sterdyniak_2012, Sterdyniak2015, He2015, Moller2015, He2017, Gerster_2017, Andrews2018, Dong2018, Rosson_2019, Palm2021}. At the same time, unconventional phases with sign-reversed Hall conductivity have been proposed to be stable in the weak-field limit near Mott phases with integer filling~\cite{Huber2011}. In light of the recent experimental progress, further understanding of these phases with anomalous Hall response is therefore a pertinent challenge.

\begin{figure}
	\centering
\includegraphics[width=\columnwidth]{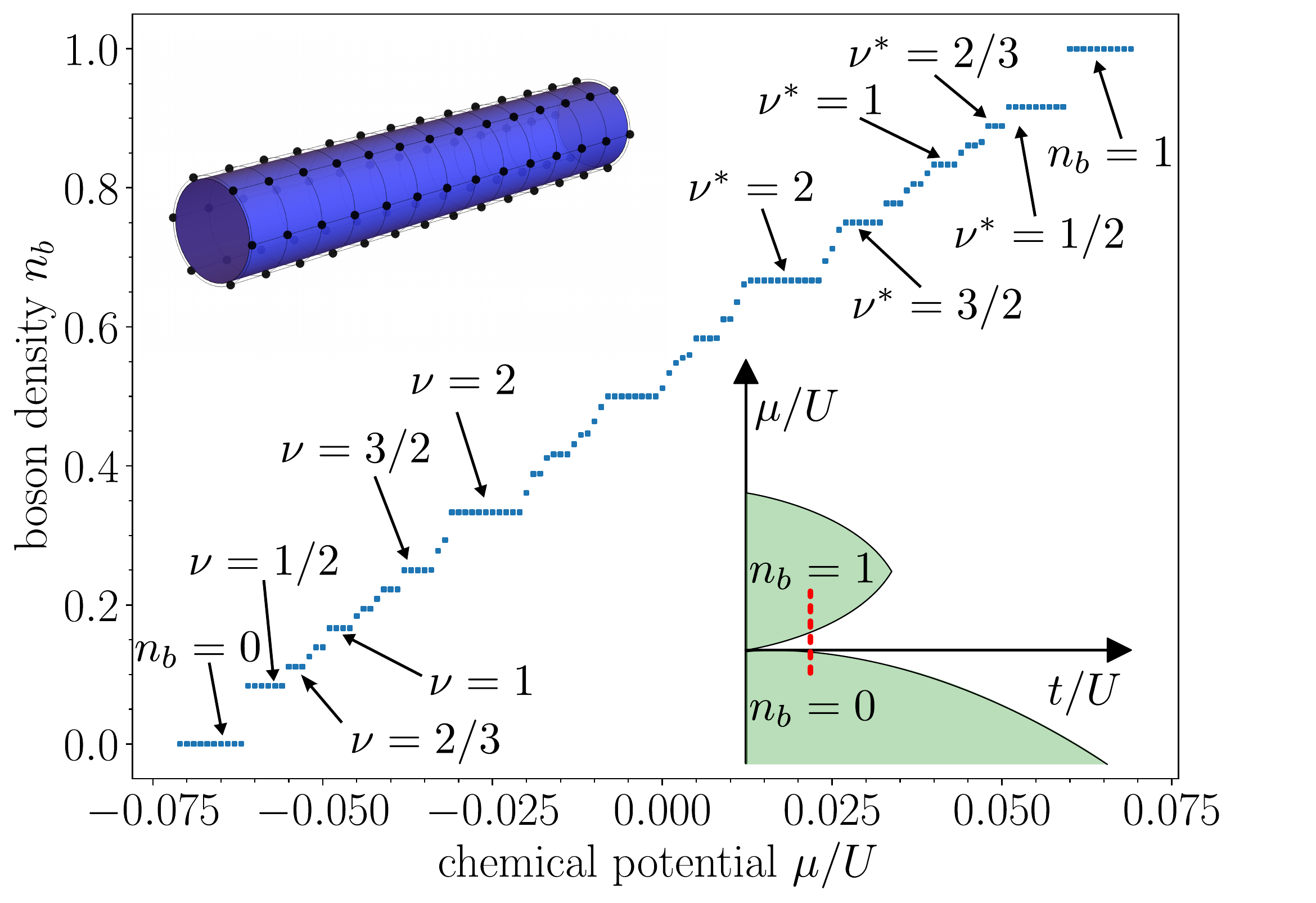}
	\caption{\textbf{Fractional quantum Hall phases of particles and holes.} The average boson density $n_b$ of the ground state as a function of the chemical potential $\mu/U$ for strong interactions $U/t = 50$ and flux $\phi = \pi/3$, obtained on an infinite cylinder with circumference $L_y = 6$ (inset, top left). We find gapped incompressible phases at various relevant filling factors. The plateaux labelled here correspond to the lowest states in the Jain sequence $\frac{p}{p + 1}$ and the Read-Rezayi sequence $\frac{p}{2}$ with integers $p$.
	The pronounced symmetry around half filling suggests an approximate particle-hole symmetry emerging at strong interactions, with holes taking the place of particles in the wave function. We mark the hole filling factor with a star. Inset, bottom right: A sketch of the phase diagram indicating the $\mu$ cut of the main panel (red dashed line). }
	\label{fig:MuScan}
\end{figure}

In this work, we numerically study fractionalized quantum Hall phases stabilized by interacting bosons on a lattice in an effective magnetic field. While most of the existing studies focus on the limit of low filling, here, we also consider fillings close to the first Mott state which has unit filling. Using the infinite Density Matrix Renormalization Group (DMRG) \cite{White1992,Hauschild_2018} algorithm on cylinder geometries we obtain not only information about the phase diagram arising from the interplay of lattice effects and magnetic flux, Fig.~\ref{fig:MuScan}, but also about key signatures of the phases. The DMRG simulations allow us to directly characterize the topological properties by calculating the Hall response (by flux threading) \cite{Laughlin1981}, the topological entanglement entropy \cite{Kitaev2006,LevinWen2006}, and the entanglement spectrum \cite{Li2008}. Besides finding different topological phases of particles near the vacuum state, we also find topological phases where holes take the place of particles near the first Mott lobe. These phases possess a sign-reversed Hall conductivity compared to their particle counterparts, providing a direct explanation for the mechanism proposed in Ref. \onlinecite{Huber2011}.

This work is structured as follows: In Sec. \ref{sec:model} we introduce the model and discuss the structure of the phase diagram. In Sec.~\ref{sec:characterization}, we characterize the fractional quantum Hall phases. We close by providing a summary and an outlook in Sec.~\ref{sec:outlook}.

\section{Model and Phase Diagram \label{sec:model}}

We consider the Bose-Hubbard model on a square lattice subjected to an effective magnetic field, resulting in a magnetic flux through each of the plaquettes of the lattice.
Our Hamiltonian is given by:

\begin{equation}
    H = -t\sum_{\langle i,j \rangle} (e^{iA_{ij}} a^{\dagger}_i a_j + \text{h.c.}) + \frac{U}{2}\sum_{i} n_i (n_i -1) - \mu \sum_{i} n_i,
    \label{eq:Hamiltonian}
\end{equation}
where $a_i^{(\dagger)}$ annihilates (creates) a boson at site $i$ and $n_i = a_i^\dagger a_i$ is the boson number operator at site $i$. The first term describes nearest-neighbor hopping with an amplitude $t$ and a background flux of $\sum A_{ij} = \phi = 2\pi n_\phi$ on each plaquette. Unless otherwise stated, we focus on the case $\phi = \pi/3$ ($n_\phi = 1/6$); other flux values give qualitatively similar results (the number of stable gapped phases we describe in the following increases with lower $n_\phi$). The second and third term describe the repulsive on-site interactions with strength $U$ and the chemical potential $\mu$ fixing the total particle number, respectively. In our numerical implementation, we assume periodic boundary conditions in the finite $y$-direction and open boundary conditions otherwise, thus realizing a cylindrical geometry, see inset of Fig. \ref{fig:MuScan} for an illustration. For high interaction strength $U/t =50$, we cut off our local Hilbert space (which is formally infinite-dimensional) at a maximum of two bosons per site, as higher occupation numbers do not change the numerical results. For interactions weaker than $U/t = 50$, we include states with three bosons per site.

For certain filling factors $\nu = \frac{n_b}{n_\phi}$, where $n_b$ is the number of bosons per site, we may expect bosonic fractional quantum Hall phases. Building on the well-established Laughlin state at $\nu = 1/2$ \cite{Laughlin1983}, one can derive certain sequences of quantum Hall states which are predicted at other filling factors. One of these is the \emph{Jain} sequence, in which flux quanta create so-called composite fermions by binding to the lattice particles \cite{Jain1989, Moller2009}. These new particles themselves form an integer quantum Hall state filling $p$ Landau levels. For the attachment of a single vortex per boson, one obtains $\nu = \frac{p}{p + 1}$, $p =1, \pm 2, \pm 3, \dots$ for the original particles. Of particular interest is the state with $p = -2$ at $\nu = 2$, the so-called Bosonic Integer Quantum Hall (BIQH) state \cite{Senthil2013, Furukawa2013, Sterdyniak2015, He2015, Moller2015}. This is an example of a symmetry-protected topological (SPT) phase, which contrasts with other fractional quantum Hall states that have intrinsic topological order.
\begin{figure}
	\centering
\includegraphics[width=\columnwidth]{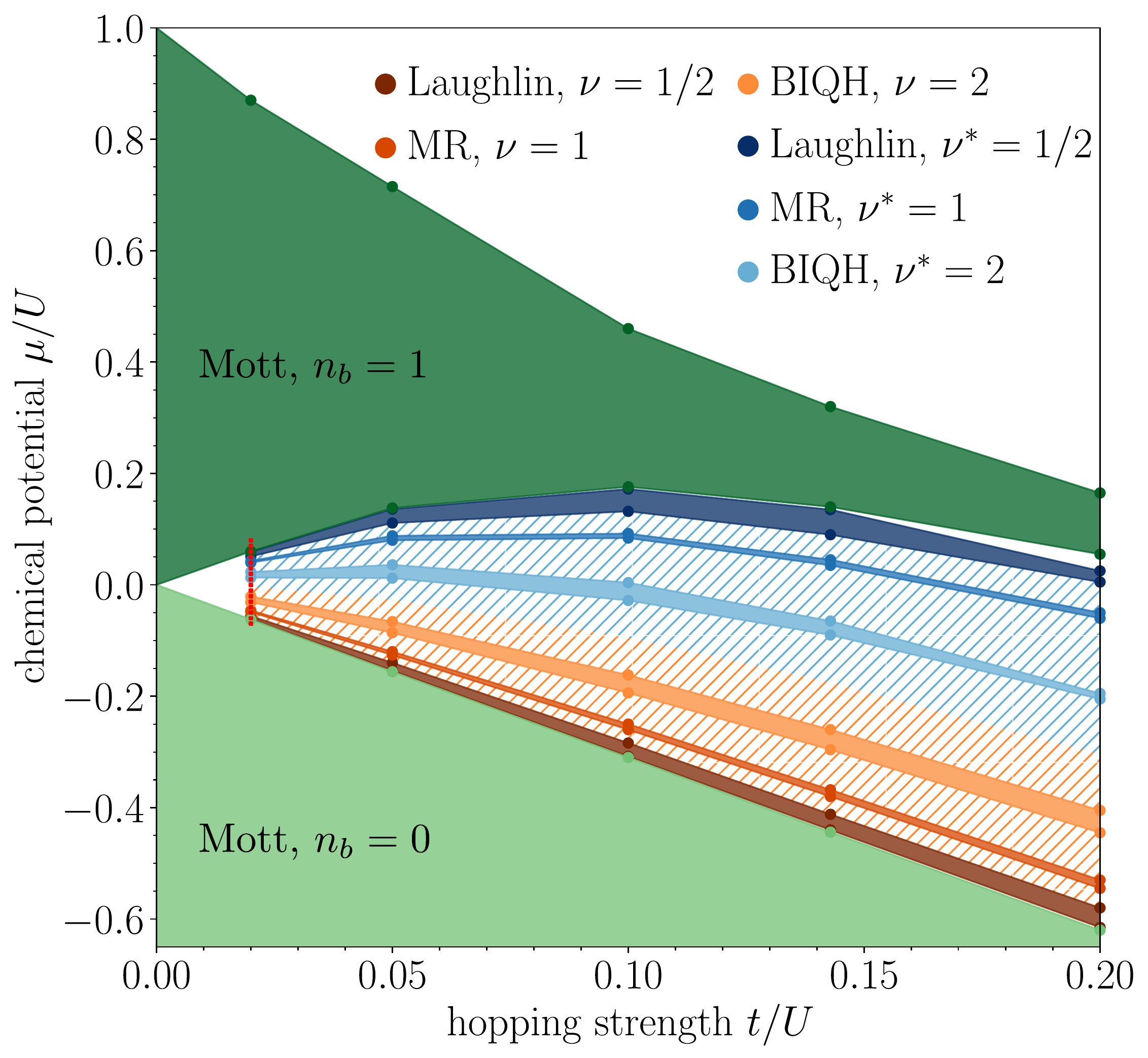}
	\caption{\textbf{Selected incompressible phases.} Phase diagram of the Bose-Hubbard model with a magnetic flux $\phi = \pi/3$ through each plaquette on an infinite cylinder with circumference $L_y = 6$, focused on the area between the vacuum and the first Mott lobe. Specific gapped states at hole filling factors $\nu^{(*)} = 1/2,1,2$ are tracked as a function of the hopping strength $t/U$. 
	We expect further gapped states (indicated by the hatched areas) corresponding e.g. to the Jain and Read-Rezayi sequence states and their hole analogues. To obtain the phase diagram, we perform infinite-size DMRG calculations, which overestimate the size of the trivial Mott state as well as the non-trivial gapped states due to the finite size in $y$ direction (see main text for details). The dotted line at $U/t = 50$ and around $\mu/U = 0$ indicates the cut considered in Fig. \ref{fig:MuScan}.}
	\label{fig:PhaseDiag}
	\end{figure}
	
\begin{table*}
\centering
\begin{tabular}{l  c  c  c  c}
\hline
\hline
\qquad State & Filling factor $\nu^{(*)}$ & Hall conductance $\lvert \sigma_{xy}\rvert$ & Anyonic statistics &  \qquad TEE $\gamma$ \qquad \qquad   \\
\hline
\qquad Laughlin & $1/2$ & $\frac{1}{2} \frac{e^2}{h}$ & 2 abelian anyons & $\log{\sqrt{2}}$ \\ \\
\qquad Bosonic Integer (BIQH) & 2 & $2 \frac{e^2}{h}$ & no fractional statistics  & 0  \\ \\
\qquad Moore-Read (MR) & 1 & $\frac{e^2}{h}$ & Ising anyons & $\log{2}$\\ \\
\hline
\hline
\end{tabular}
\caption{\textbf{Properties of selected quantum Hall states.} The theoretically predicted absolute values of the Hall conductance and the filling factor are proportional. The Laughlin state hosts two abelian anyons with semionic statistics, leading to a value of the topological entanglement entropy (TEE) $\gamma = \log \sqrt 2$. The BIQH state is a symmetry-protected topological phase and therefore lacks fractional statistics and TEE. The Moore-Read state is connected to the Ising conformal field theory, so its anyonic content consists of the identity, a fermion and an Ising anyon, yielding a TEE $\gamma =\log 2$.}
\label{tab:States}
\end{table*}

Another family of quantum Hall states is predicted by the \emph{Read-Rezayi} sequence, which are candidates for the ground states at fillings $\nu = \frac{p}{2}$, $p =1, 2, 3, \dots$. Except for the abelian Laughlin state at $p = 1$, these are non-abelian topological states with more complex behavior; among them is the Moore-Read state with $p = 2$ at $\nu = 1$ which hosts an Ising set of anyons \cite{Moore1991}. In the continuum limit, these trial states can be seen as the densest ground states of a repulsive $p+1$-body interaction \cite{Read1999, Cooper2008, Cooper2001}. Several studies have found promising signs that a number of these phases are actually realized in different lattice models; in particular, the Laughlin, Jain and Moore-Read states have been suggested as ground states for their respective filling factors in the system we consider here. Nevertheless, reliably predicting the actual state at a given configuration in the thermodynamic limit remains a challenging task, since different quantum Hall states may compete with other states such as charge density waves and many ground states thus elude a simple characterization. 

A first step to obtain some information about the different phases is to calculate the filling factor as a function of the chemical potential $\mu$. This is shown in Fig. \ref{fig:MuScan} for the strongly repulsive case $U/t = 50$. The average particle number is obtained by infinite DMRG on a $L_y = 6$ cylinder in the grand canonical ensemble using the TeNPy package~\cite{Hauschild_2018}. This algorithm provides us with an approximate ground state of the infinite system expressed as matrix-product state with a bond dimension of $\chi$. 
From this representation we can directly calculate the average particle number as function of the chemical potential. We find an ample number of plateaux which suggest gapped, incompressible phases at the filling factors predicted by the Jain and Read-Rezayi sequences. Interestingly, we also observe a family of plateaux at higher filling factors $n_b>1/2$. Introducing a complementary hole filling factor $\nu^* = \frac{1- n_b}{n_\phi}$ (marked with $*$ throughout) we observe an approximate symmetry between plateaux at the same particle filling $\nu$ and hole filling $\nu^{*}$. 

Starting from a semiclassical picture of vortices in bosonic systems, Huber and Lindner postulated the existence of regions below the Mott lobes characterized by sign-reversed Hall conductivity for sufficiently strong interactions~\cite{Huber2011}. One can define a proportionality factor $\alpha$ between the Berry phase resulting from the movement of a vortex in a closed loop and the area enclosed by this loop. The Hall conductivity $\sigma_{xy}$ is proportional to this factor, $\sigma_{xy} \propto \alpha$. In systems exhibiting Galilean invariance, this factor can be indentified with the number of superfluid bosons~\cite{Haldane1985,Ao1993}. In lattice systems, which break Galilean symmetry, the commutation relations of vortex translation operators define the Berry phase when moving around a single plaquette only up to an integer, so $\sigma_{xy} \propto \alpha = \left(n_b +k\right)$, $k \in \mathbb{Z}$. By deriving effective theories from the Bose-Hubbard model in a weak magnetic field, one can show that this integer jumps on lines which emanate from particle-hole symmetric points, including ones which lie between two Mott lobes in the hard-core limit $U/t\to \infty$, thus allowing for negative Hall conductivity below the Mott lobes. This directly connects the two types of filling factors $\nu \leftrightarrow \nu^*$ for $U/t \rightarrow \infty$. In a similar vein, some numerical studies suggest the emergence of a Laughlin state of holes on the background of the $n_b =1$ Mott insulator \cite{Umucalilar2007, Umucalilar2010, Kuno2017, Natu2016}. Combining these ideas, we argue that, for strong interactions, the plateaux above $n_b = 1/2$ can be seen as particle-hole analogues of the respective gapped phases at low filling factors. Replacing particles with holes directly explains a reversed Hall conductivity in this region.

In Figure \ref{fig:PhaseDiag}, we show a phase diagram
of our model obtained for an infinite cylinder with circumference $L_y = 6$, with the same method as for Fig~\ref{fig:MuScan}.
We focus on the area between the vacuum and the first Mott lobe, where we expect quantum Hall phases of particles and holes and track states with three specific filling factors $\nu^{(*)} = 1/2, 1, 2$ which may host states in the Jain and Read-Rezayi sequences; orange and blue shaded regions. Around these phases additional incompressible phases exist, which we have not extracted in full detail (indicated by the hatched areas).
While the simulations so far allow us to make some general predictions about the system, they also suffer from strong finite-size effects because the considered cylinder circumference is comparatively small, compared to the magnetic unit cell. This leads to an overestimation of the stability of the gapped phases. The phase boundaries should therefore be regarded only as guides.

\section{Characterization of quantum Hall phases \label{sec:characterization}}
The calculated observables so far give insight into the emergence of new phases due to magnetic flux in the Bose-Hubbard model. However, they do not offer more specific information about which states are actually realized. In the following, we will discuss several properties which allow us to classify the ground states. We focus on the $\nu^{*} = 1/2, 2$ states and the corresponding particle states $\nu = 1/2, 2$ to study the supposedly symmetric behavior. For these states, as well as for the trivial Mott insulator with $n_b=1$, we consider strong but finite interactions $U/t = 50$. At $\nu = 1$, a gapped state has been found which has been predicted to be a Moore-Read state in Ref. \onlinecite{Palm2021}. In the continuum, the Moore-Read state is a zero-energy state of a repulsive three-body interaction~\cite{Cooper2008}. Therefore, for this state we consider comparatively weak interactions $U/t = 2$, for which we expect two particles occupying the same site not to be disfavored too strongly, whereas three-particle occupations are still rather strongly suppressed. This regime favors a Moore-Read state as a possible ground state and renders its topological properties more tractable. 
In general, finding a direct particle-hole conjugate of a state with two holes occupying the same site cannot occur below the first Mott lobe (but can arise between higher Mott lobes), yet, different lattice regulated variants of the Moore-Read state of holes are conceivable. 
In the case of moderate interactions $U/t = 20$, we have indeed found well-converged results yielding a quantized Hall conductance as predicted by the Moore-Read theory. However, further studies of the robustness of this state in the thermodynamic limit are required to fully clarify the situation.
For reference, we summarize in Table \ref{tab:States} the most important characteristics of the candidate states at these filling factors~\footnote{Another candidate for $\nu^{(*)} = 2$ would be the $p = 4$ Read-Rezayi state, which we have not listed}.

\subsection{Hall Conductivity}

One of the most direct signatures of quantum Hall states with particle number conservation is their quantized \emph{Hall conductance} $\sigma_{xy}$. 
To compute the Hall response, we can follow Laughlin's argument and adiabatically thread a flux quantum $\Phi_y = 2\pi$ though the cylinder while monitoring how much charge is transferred across a cut through the cylinder. To this end, we make use of the Schmidt decomposition of the ground state $\ket{\Psi} = \sum_{i=1}^\chi \Lambda_{i} \ket{i_L} \otimes \ket{i_R}$. As the Hamiltonian conserves the $U(1)$ particle number, we associate a charge value $Q^{L}_{i} \in \mathbb{Z}$ with every Schmidt state $\ket{i_{L}}$. The average charge across the cut is obtained from the Schmidt representation as $\langle Q^{L}(\Phi_y) \rangle = \sum _{i} \Lambda_{i}^2(\Phi_y) Q^{L}_{i}(\Phi_y)$. Measuring this charge while threading a flux quantum $\Phi_y = 2\pi$ through the cylinder gives the Hall conductance~\cite{Zaletel2014, Grushin2015}

\begin{equation}
    \sigma_{xy} = \frac{e^2}{h} [\langle Q^{L}(\Phi_y=2\pi)  \rangle - \langle Q^{L}(\Phi_y=0)  \rangle].
    \label{eq:HallCond}
\end{equation}
The pumped charge is quantized to an integer or a fractional value depending on the specific quantum Hall state and filling factor, as given in Table \ref{tab:States}.

\begin{figure}
	\centering
\includegraphics[width=\columnwidth]{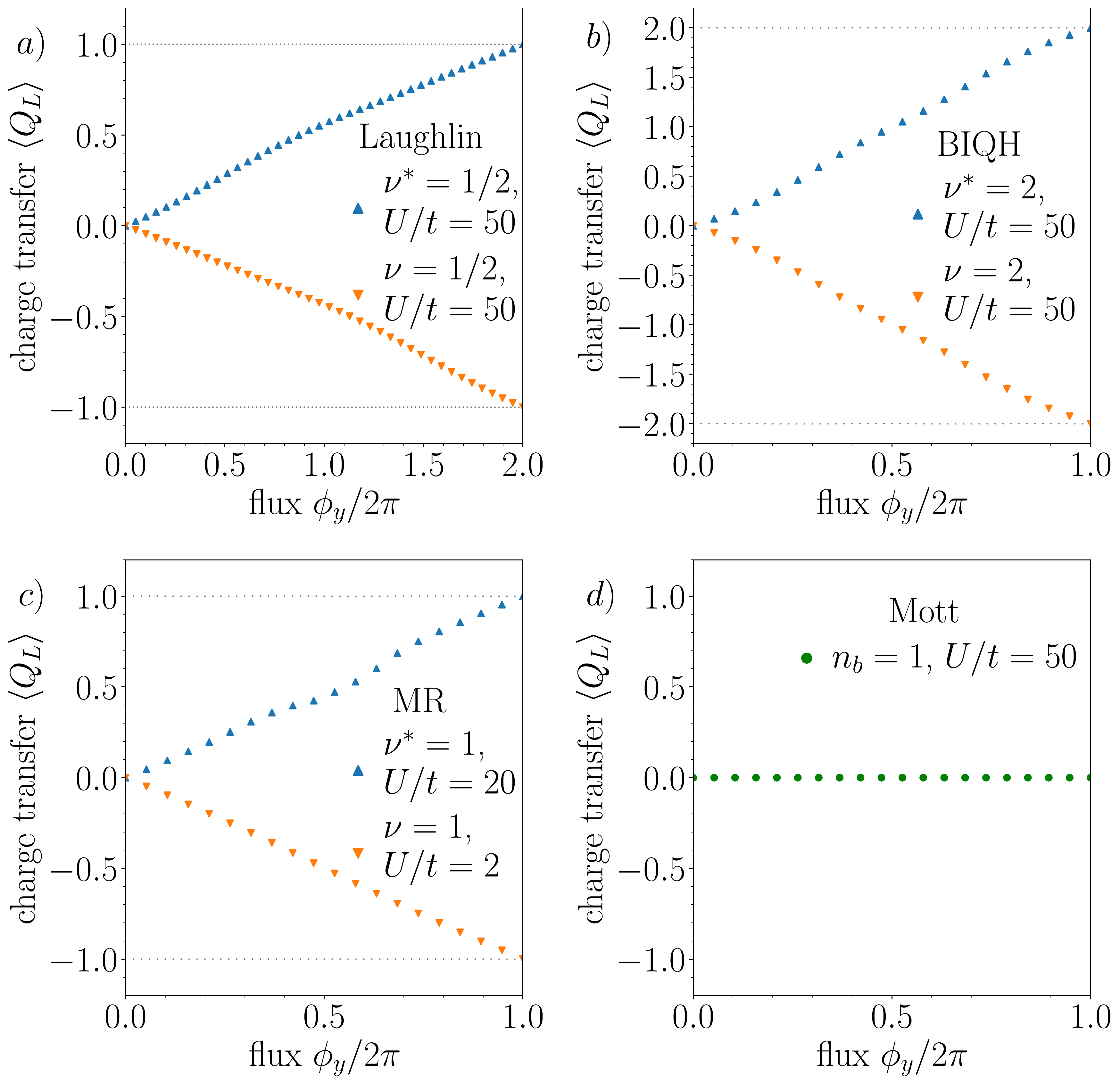}
	\caption{\textbf{Charge pumping.} a) Charge pumping for the ($\nu^{(*)} = 1/2$)-states for two periods. A single charge is transferred from one half to the other, as expected for the Laughlin state. The sign reversal between both states suggests an approximate particle-hole symmetry of the phases. The other states are analyzed over one pumping period: In b), two charges are transferred per flux period for the ($\nu^{(*)} = 2$)-states, again with different sign between the states. In c), a single charge pumped is indicative of the Moore-Read state at $\nu^{(*)} =1$ at different interaction strengths. By contrast, the trivial Mott insulator in d) does not allow any pumping, highlighting the difference between a trivial insulator and the more intricate topological states.}
	\label{fig:FluxThread}
	\end{figure}

In Fig. \ref{fig:FluxThread}, we show $\langle Q^{L}(\Phi_y) \rangle$ for various gapped states. The cylinder considered has a circumference of $L_y = 6$, and we fix the bond dimension to $\chi = 900$. We repeatedly calculate the ground state of the system at a certain $\Phi_y$ before slightly increasing the gauge flux, thereby simulating an adiabatic tuning of an electromotive force alongside the $y$ axis. This induces charge pumping along the $x$ direction.

For the ($\nu^{(*)} = 1/2$)-states shown in Fig.~\ref{fig:FluxThread}(a), we increase our flux value up to $\Phi_y = 4 \pi$. For a Laughlin-type state at this filling with a doubly degenerate ground state, this procedure will let the ground state at $\Phi_y = 0$ flow back to itself at $\Phi_y = 4\pi$. Accordingly, we obtain an almost perfect quantized value of a single charge being transferred in two pumping periods, leading to half-a-charge per flux quantum and a Hall conductance of $\lvert \sigma_{xy}\rvert = \frac12 \frac{e^2}{h}$ as expected for a $\nu = 1/2$ Laughlin state. We also find that between the two states at particle-hole symmetric filling the sign of the Hall conductance is reversed, as our argument predicted, \textit{c.f.} Ref. \cite{Huber2011}.

We find a similar reversal of the Hall conductance for the ($\nu^{(*)} = 2$)-states [see Fig.~\ref{fig:FluxThread}(b)]. The Hall conductance of $\lvert \sigma_{xy}\rvert = 2 \frac{e^2}{h}$ suggests a BIQH state. At $\nu = 1$ for interaction strength $U/t = 2$ and at $\nu^* = 1$ for $U/t = 20$ [Fig.~\ref{fig:FluxThread}(c)] we find a conductance quantized to unity $\lvert \sigma_{xy}\rvert = \frac{e^2}{h}$ after one period, which is a characteristic property of the Moore-Read state not yet addressed in the existing literature. Here too, the characteristic sign-reversal is present.

In contrast to the previously discussed states, the trivial Mott insulator shown in Fig.~\ref{fig:FluxThread}(d) does not exhibit any charge pumping. These results establish their topological nature, as well as the sign change in the Hall conductivity when going from particles to holes.

\subsection{Topological Entanglement Entropy}

\begin{figure*} 
	\centering
\includegraphics[width=2\columnwidth]{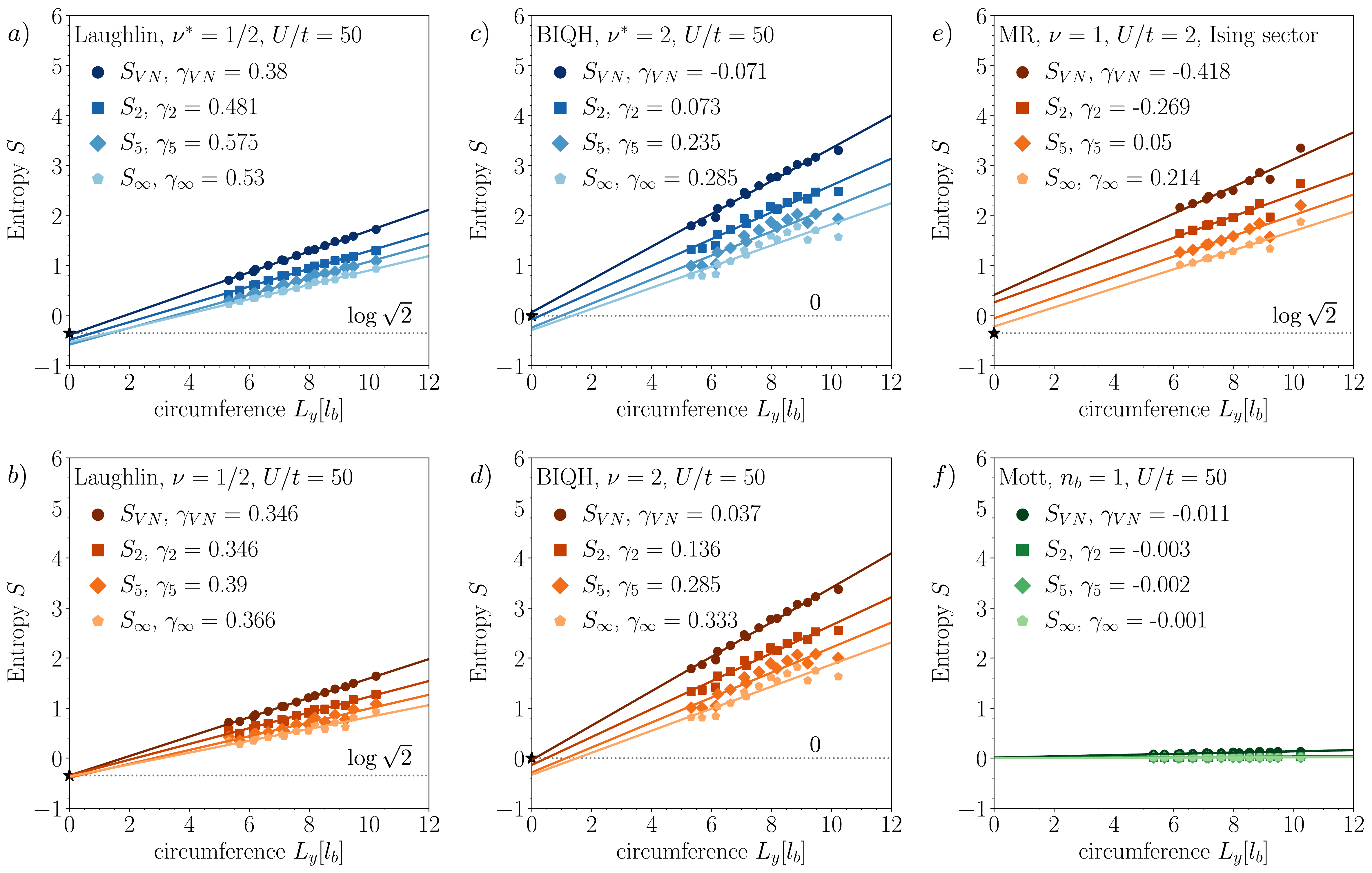}
	\caption{\textbf{Finite-size flow of R\'enyi entropies}. (a) and (b) Extrapolation of R\'enyi entropies for $\nu^{(*)} = 1/2$  gives strong indications of Laughlin-type states at these filling factors. (c) and (d) States with $\nu^{(*)} = 2$ are more difficult to interpret due to stronger deviations between the different entropy measures. However, the extrapolations of the low-order entropies are consistent with zero as expected for the Bosonic Integer Quantum Hall state. The values for the Ising sector of the Moore-Read state in (e) suffer from the same problem, high-order entropies however are relatively close to the expected value. (f) Comparison for a Mott insulator, which yields zero topological entanglement entropy. Different flux densities $n_\phi = 1/6, 1/7, 1/8$ are analyzed, so the cylinder circumference $L_y$ is expressed in terms of the respective magnetic length $\ell_B$.}
	\label{fig:TEE}
	\end{figure*}

Entanglement measures carry relevant information about the topological properties of the state and can be directly obtained from the Schmidt values. A particularly important measure is the von-Neumann entanglement entropy $S_\text{vN} =  -\sum_{i} \Lambda_{i}^2 \log{\Lambda_{i}^2}$, which is a special limiting case of the more general R\'enyi entropies defined by $S_{\alpha} = \frac{1}{1-\alpha} \log{\left(\sum_{i} \Lambda_{i}^{2\alpha}\right)}$. For $\alpha \rightarrow \infty$, we obtain the infinite R\'enyi (max) entropy $S_{\infty} = -\log{\left(\max{\left[\Lambda_{i}^2\right]}\right)}$.  
For cylinder geometries of circumference $L_y$, all R\'enyi entropies for minimally-entangled ground states \cite{Dong2008,Zhang2012} corresponding to abelian anyons obey a modified area law of the form
\begin{equation}
    S_{\alpha} = c_{\alpha} L_y - \gamma + \mathcal{O}(1/L_y),
    \label{eq:FinSize}
\end{equation}
where the constants $c_{\alpha}$ are non-universal. However, the correction $\gamma$ to the area law is a universal quantity characterizing the topology of the state and is identical for every entropy. In finite-size systems, we can thus compute the scaling of different entropies with respect to system size to extract information about the topological order of the state. 
The universal contribution $\gamma$ is referred to as \emph{topological entanglement entropy} (TEE) \cite{Kitaev2006, LevinWen2006} and provides insight in the anyonic quasiparticle excitations. For a given ground state manifold, it is given by $\gamma =\log{\left(\sqrt{\sum_i d_i^2}\right)}$, where $d_i$ is the quantum dimension of the anyon associated to the topological sector $i$. 

Laughlin states at $\nu = 1/m$ have a particularly simple topological entanglement entropy because they host only $m$ different abelian anyons (i.e. their exchange statistics can be described with a complex phase factor $e^{i\pi/m}$) for which $d_i = 1$.  
A Laughlin state at half-filling with $m=2$ thus has $\gamma = \log\sqrt{2} \approx 0.347$. 
The Moore-Read state has a richer anyon content and includes in addition to the identity two non-trivial anyons: a fermion and an Ising anyon. The former is an abelian anyon with $d_i = 1$ and the latter is non-abelian with $d_i = \sqrt{2}$. Therefore, the ground state manifold of the Moore-Read state has $\gamma = \log{2} \approx 0.693$ \cite{Zhu2015}. For this state, a further correction of the topological entanglement entropy is present in the cylinder geometry considered here arising from its non-abelian statistics. In particular, for the ground state in the topological sector $i$, the constant $\gamma$ in Eq.~\ref{eq:FinSize} is given by $\gamma_i = \gamma - \log d_i$~\cite{Dong2008, Zhang2012, Cincio2013, Zaletel2013}. While this does not affect the vacuum or fermion sectors where $d_i = 1$, the Ising sector with $d_i = \sqrt{2}$ should extrapolate to $\gamma_i = \log \sqrt{2}$. The BIQH state does not have topological ground state degeneracy; rather, it is a symmetry-protected topological phase and satisfies $\gamma = 0$.

\begin{figure*} 
	\centering
\includegraphics[width=2\columnwidth]{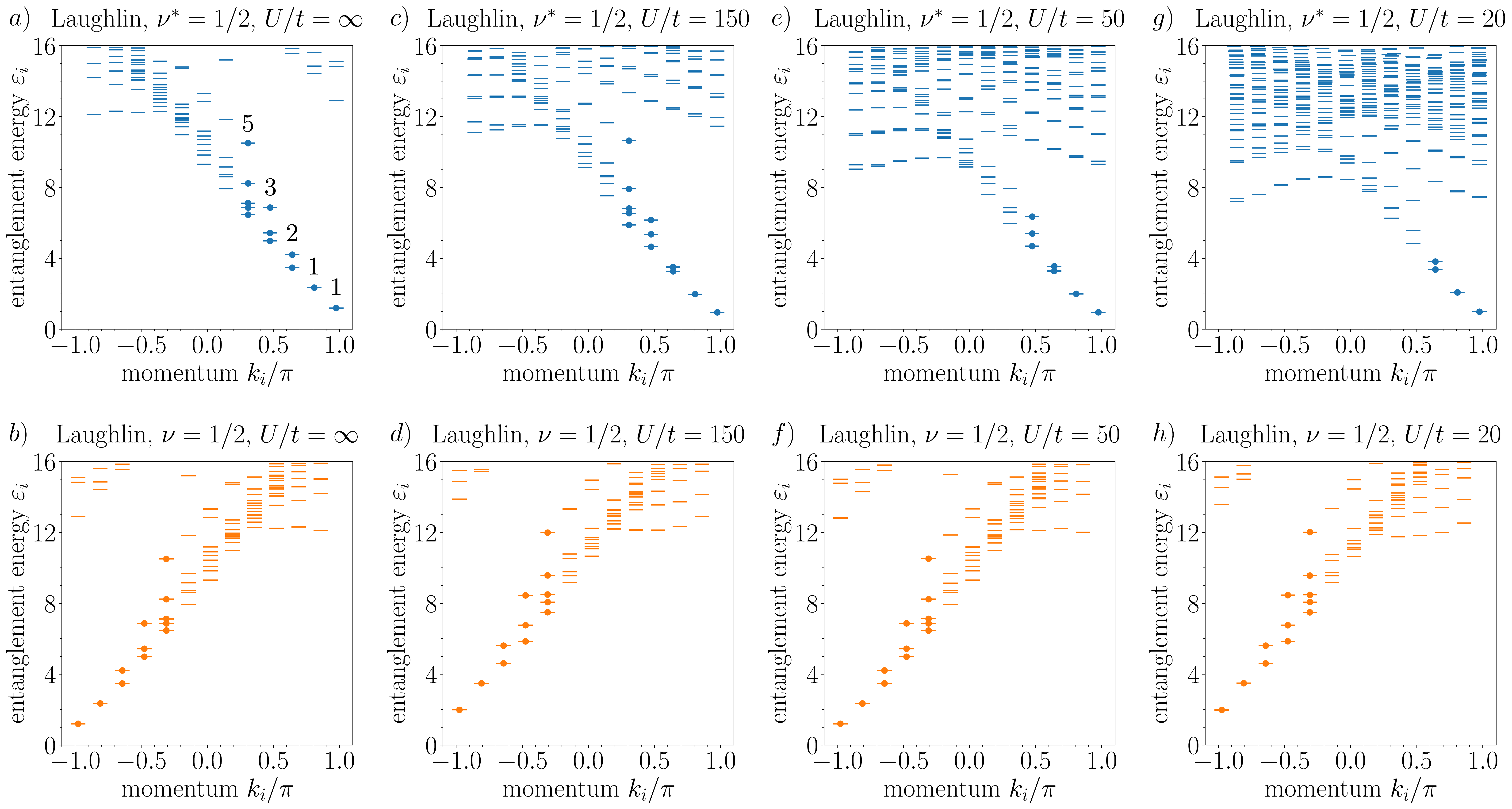}
	\caption{\textbf{Entanglement spectra of the Laughlin states.} The entanglement spectrum $\varepsilon_{i}$ as a function of the momentum in the $y$ direction $k_i$ for the $\nu^{(*)} = 1/2$ Laughlin states at different interaction strengths $U/t$ on a $L_y = 12$ cylinder. In the upper row, we show the spectrum for the $\nu^* = 1/2$ hole state. While the hard-core limit clearly obtains the counting sequence $\{1,1,2,3,5\dots\}$ predicted by the conformal field theory of the Laughlin state (indicated by circles in the images), smaller values for $U/t$ lead to a background of low-lying Schmidt states that obscure this sequence. By contrast, the spectra for the particle state at $\nu = 1/2$ are almost unaffected by weaker interactions due to the lower filling of bosons. Note that the entanglement spectra direction differs between particles and holes, which is consistent with the sign reversal of the Hall conductivity. For reasons of clarity, only the $Q_{i}^L = 0$ charge sector is shown. Other charge sectors exhibit the same counting.}
	\label{fig:EntSpect}
	\end{figure*} 

\begin{figure*} 
	\centering
\includegraphics[width=2\columnwidth]{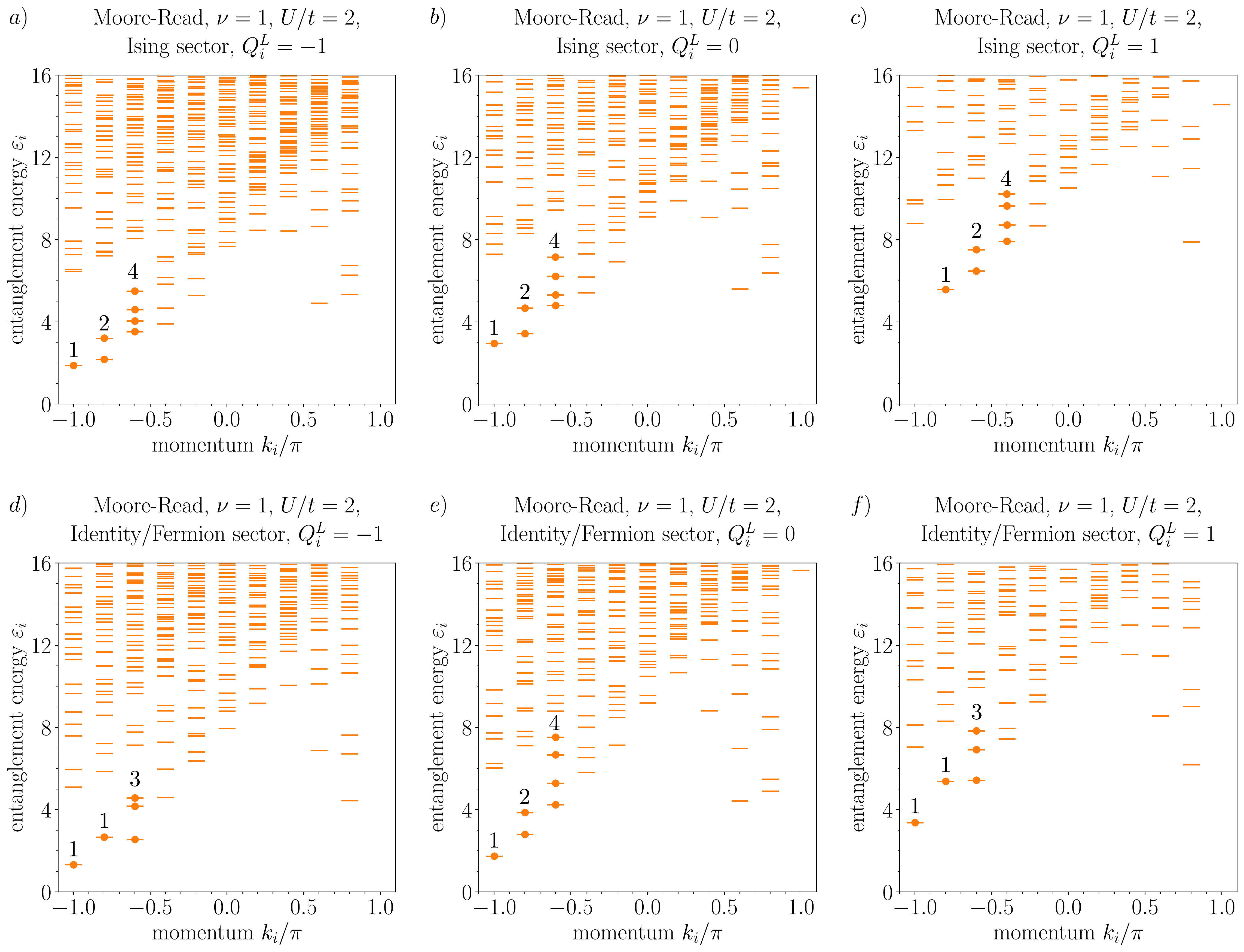}
	\caption{\textbf{Entanglement spectra of the Moore-Read state.} The entanglement spectrum $\varepsilon_{i}$ as a function of the momentum in the $y$ direction $k_i$ for the $\nu = 1$ Moore-Read state at different flux densities $n_\phi$ on a $L_y = 10$ cylinder with interaction strength $U/t = 2$, resolved by their charge value. In the upper row, we show the spectrum for the state at $n_\phi = 1/6$, as obtained by infinite DMRG. It clearly follows the counting $\{1,2,4\dots\}$ in every charge sector, which is characteristic of the Ising topological sector as given by the Moore-Read conformal field theory. In the lower row, we show the entanglement spectra obtained for $n_\phi = 1/7$. Here, the $\{1,2,4\dots\}$ counting is present only in the even charge sectors, while the odd sectors follow the sequence $\{1,1,3\dots\}$ in the low-energy space. This alternating behavior is characteristic of the other two sectors, the Identity sector and the Fermion sector.}
	\label{fig:EntSpectMR}
\end{figure*} 

Here, we use this expected scaling behavior to obtain information about the (non)-topological order for the ground states of the relevant filling factors. We calculate four different orders of R\'enyi entropies (including the von-Neumann entropy and the max-entropy) for the circumferences values $L_y = 6, 7,8,9,10$ and the flux densities $n_\phi = 1/6, 1/7, 1/8$, keeping the filling fraction fixed, and extrapolate to $L_y = 0$ to estimate the topological entanglement entropy $\gamma$. Varying the flux density allows us to access many more data points than for a fixed flux density. When performing the extrapolation, we express the circumference $L_y$ in units of the magnetic length $\ell_B = \left(2\pi n_\phi \right)^{-1/2}$~\cite{Bauer2016, Andrews2018, schoonderwoerd2019interactiondriven}. We use maximal bond dimensions of $\chi = 2500$ for all states.

Our results for the entanglement entropies are shown in Fig.~\ref{fig:TEE}. For the Laughlin states at $\nu^{(*)}=1/2$ the extrapolation of the finite-size entropies are well converged to the expected value, Fig.~\ref{fig:TEE}(a) and (b). This agreement further confirms our hypothesis of a Laughlin state for this hole filling factor. The values for the ($\nu^{(*)} = 2$)-state in  Fig.~\ref{fig:TEE}(c) and (d) are more difficult to interpret, as we find the extrapolated value for $\gamma$ varies more strongly with the R\'enyi entropy order. While the extrapolation of the von-Neumann entropies give values which are close to zero as predicted by the BIQH theory, other orders yield higher values. Further investigation into the actual structure of these states is thus necessary, particularly because other studies have demonstrated properties at $\nu = 2$ which are sometimes inconsistent with the BIQH state \cite{Andrews2018,Regnault2003}. However, the similarities between these two states suggest that they are indeed particle-hole conjugates.

In Fig~\ref{fig:TEE}(e), the results for the supposed Moore-Read state at $\nu = 1$ are shown. Here, we have chosen only those values whose entanglement spectra suggest that the ground state in the Ising sector is realized, which should lead to a TEE of $\gamma_{i} = \log \sqrt{2}$. (We show the differences in the entanglement spectra in the next section.) We are therefore left with fewer data points than in the other cases. Here too, we note significant differences between the numerical values, which could arise due to finite-size effects, or an insufficient number of data points. However, the infinite R\'enyi entropy $S_{\infty}$ curve leads to a value which is rather close to $\gamma_{i} = \log\sqrt{2}$. Preparing the different topological sectors could further help to  to fully characterize this state.

For reference, we show in Fig.~\ref{fig:TEE}(f) the entanglement scaling for the Mott insulator. All data points are close to zero, as expected for a state which is close to a product state.

\subsection{Entanglement Spectra}
In this section, we focus on the Laughlin states at $\nu^{(*)} = 1/2$ and the Moore-Read state at $\nu = 1$. We wish to characterize these further by computing their \emph{entanglement spectra} \cite{Li2008}. In the cylinder geometry, translational symmetry in the $y$ direction (for the right choice of magnetic unit cell) ensures that the Schmidt states are momentum eigenstates in the $y$ direction. Therefore, we can label each Schmidt value and each \emph{entanglement energy} $\varepsilon_{i} = - 2 \log \Lambda_{i}$ with a $k_i$ value. For quantum Hall states, the lowest lying levels of these relationship give insight into the edge conformal field theory governing the state. For the Laughlin state, one excepts a typical counting  $\{1,1,2,3,5\dots\}$ of a $U(1)$ chiral boson for each charge sector $Q^L$ \cite{Cincio2013, Grushin2015}. For the Moore-Read state, the low-energy behavior is more complicated, as its conformal field theory also hosts a Majorana fermion besides the chiral boson. Therefore, the entanglement spectrum depends on the topological sector. In the identity sector, the counting depends on the charge sector $Q_i^L$: Alternating with $Q_i^L$, it is either $\{1,1,3,5,10\dots\}$ or $\{1,2,4,7\dots\}$. For the fermion, the low-lying energy levels follow the same structure shifted by $\Delta Q_i^L = 1$. The Ising sector follows a universal $\{1,2,4,8\dots\}$ counting in every charge sector, thus it is easily distinguishable from the other two sectors.~\cite{Liu2012, Zhu2015}

We compute the entanglement spectra for the two Laughlin states with $\nu^{(*)} = 1/2$ as a function of the interaction strength $U/t$; Fig.~\ref{fig:EntSpect}. The entanglement spectra are shown only for the $Q_{i}^L = 0$ charge sector, for reasons of clarity; other charge sectors exhibit the same structure shifted in $k_i$, as expected from the theoretical prediction. We calculate the ground state with a bond dimension of at least $\chi = 1800$ for the four different interaction strengths $U/t = \infty, 150, 50, 20$ on a $L_y = 12$ cylinder. Accordingly, we obtain twelve possible values for the quantum number $k_i$. As indicated by the dots in Fig.~\ref{fig:EntSpect}, the characteristic counting for the conformal field theory corresponding to the Laughlin state is clearly visible; particularly at strong interactions. We also observe that the sequence reverses for the two cases, which indicates the particle-hole correspondence. Moreover, we note that the particle-hole symmetry reflected in the entanglement spectra plots is broken by weak interactions away from the hard-core limit. In Fig~\ref{fig:EntSpect}, the sequence counts in the bottom row (corresponding to the particle states) are virtually unchanged as $U/t$ decreases; in contrast, the hole states develop a background at higher entanglement energies which affects the sequences at weak interactions. We see that for $U/t = 20$, only the lowest three levels of the counting are still distinguishable. This is because lower interactions lead to a stronger contribution from states which have multiple bosons occupying the same site. This contrasts with the Laughlin wavefunction which in the continuum is the ground state of a Hamiltonian exactly suppressing these states \cite{Cooper2008}. Hence the Laughlin state of holes is less robust at reducing the interaction than its particle analogue.

In Fig.~\ref{fig:EntSpectMR}, we show entanglement spectra for the $\nu = 1$ Moore-Read state at $U/t = 2$ on a $L_y = 10$ cylinder with different flux densities $n_\phi$. As discussed in the preceding section, most of our simulations for the Moore-Read state have resulted in an Ising state, some however have yielded ground states from the other topological sectors, which can be distinguished by their low-energy counting. In the upper row, we show the entanglement spectrum for flux density $n_{\phi}= 1/6$, which realizes the first three values of the $\{1,2,4,8\dots\}$ counting in all three charge sectors $Q_i^L = -1,0,1$. The ground state found by DMRG in this particular configuration can therefore be identified as lying in the Ising sector. The lower row shows the spectrum for flux density $n_\phi = 1/7$, which realizes a different topological sector, as the counting for the even charge sector $Q_i^L = 0$ can be identified with the $\{1,2,4,7\dots\}$ counting, whereas the odd charge sectors $Q_i^L = \pm1$ follow the counting $\{1,1,3,5,10\dots\}$. In general, the entanglement gap is significantly smaller than for the Laughlin state at $\nu = 1/2$, as expected for the Moore-Read state which has both higher particle number and necessitates lower interaction strengths. Therefore, only a few low-energy levels are separated. Nevertheless, the typical chiral counting can still be deduced, further confirming the existence of a non-abelian topological phase in this system.

In general, our results demonstrate that the entanglement spectra are rather fragile and that the low-energy counting sequence may not be as clearly visible as expected, despite the state being topologically nontrival. 

\section{Conclusions and Outlook \label{sec:outlook}}
We have characterized gapped phases with topological order in the Bose-Hubbard model subjected to an effective magnetic field with particular focus on fillings nearby the first Mott lobe. A detailed characterization of charge pumping, topological entanglement, and the entanglement spectrum obtained numerically from matrix-product state simulations provided us with information about their topological order. We have obtained convincing data for a Laughlin state near unit filling which can be seen as the hole analogue of the $\nu = 1/2$ fractional quantum Hall state for bosons, as evidenced by the reversed sign of the Hall current. However, for other gapped states, more involved investigations are necessary to unambiguously identify their topological order. 

While we were able to identify topological properties of certain plateaux, much larger computational resources are required to obtain a complete understanding of all gapped phases that can occur. In particular, a clear understanding of the gapped states emerging at low filling factors is necessary before understanding the more unstable conjugates at higher boson density. In this regard, we have also provided additional information supporting a Moore-Read state at $\nu = 1$. For these more complex non-abelian states, protocols which distinguish the different topological ground state sectors may be able to provide deeper understanding of these phases. Further insights could also be obtained by realizing these phases experimentally with synthetic quantum systems. In particular, protocols have been developed theoretically to characterize non-trivial topological properties of such systems, including the analysis of currents~\cite{Kessler_2014, Repellin_2020, Motruk_2020, buser2021}, flux insertion~\cite{Wang2018,Raciunas2018, wang2021}, randomized measurements~\cite{Cian_2021, Kokail_2021, zache2021entanglement}, interferometry~\cite{Liang_2008, Grusdt_2016, birnkammer2020, Dehghani_2021}, circular dichroism~\cite{Tran2017,Repellin2019} and the depletion of quasi-holes~\cite{Macaluso2019, Macaluso2020}.

\textbf{Acknowledgements}. 
We thank G. M\"oller for insightful discussions. We acknowledge support from the Technical University of Munich - Institute for Advanced Study, funded by the German Excellence Initiative and the European Union FP7 under grant agreement 291763, the Deutsche Forschungsgemeinschaft (DFG, German Research Foundation) under Germany’s Excellence Strategy--EXC--2111--390814868, TRR80 and DFG grants No. KN1254/1-2, KN1254/2-1 as well as from the European Research Council (ERC) under the European Union’s Horizon 2020 research and innovation programme (grant agreement No. 851161).

\bibliography{main}

\end{document}